\documentclass[3p,12pt,preprint]{elsarticle}
\usepackage{lineno,hyperref}
\modulolinenumbers[1]

\newcommand{\bsub}{\begin{subequations}}
\newcommand{\esub}{\end{subequations}}

\newcommand{\om}{\omega}

\newcommand{\po}{\mbox{\boldmath $\omega$}}

\newcommand{\ppsi}{\mbox{\boldmath $\psi$}}

\newcommand{\pGamma}{\mbox{\boldmath $\Gamma$}}

\newcommand{\pF}{\textbf{\emph{F}}}

\newcommand{\pU}{\textbf{\emph{U}}}

\newcommand{\pe}{\textbf{\emph{e}}}

\newcommand{\pn}{\textbf{\emph{n}}}

\newcommand{\pu}{\textbf{\emph{u}}}

\newcommand{\px}{\textbf{\emph{x}}}

\newcommand{\pat}{\partial}
\newcommand{\na}{\nabla}

\newcommand{\rd}{\textrm{d}}

\newcommand{\beq}{\begin{equation}}
\newcommand{\eeq}{\end{equation}}
\newcommand{\bsubeq}{\begin{subequations}}
\newcommand{\esubeq}{\end{subequations}}
\newcommand{\beqn}{\begin{eqnarray}}
\newcommand{\eeqn}{\end{eqnarray}}
\newcommand{\fr}{\frac}
\newcommand{\lb}{\label}
\newcommand{\er}{\eqref}

\usepackage{hyperref}
\hypersetup{colorlinks,allcolors=blue}
\usepackage{amsmath}
\usepackage{multirow}
\usepackage{amsfonts,amssymb}
\usepackage{graphicx}
\usepackage{subfigure}
\usepackage[rel]{overpic}

\graphicspath{{figures/}}
\begin{document}

%\large

\begin{frontmatter}

\title{\textbf{Numerical validation and physical explanation of the universal force theory of three-dimensional steady viscous and compressible flow}}

\author[mysecondaddress]{Shufan Zou}

\author[myfirstaddress]{Luoqin Liu\corref{cor1}}
\cortext[cor1]{Corresponding author}
\ead{luoqin.liu@utwente.nl}

\author[mysecondaddress]{Jiezhi Wu}

\address[mysecondaddress]{State Key Laboratory for Turbulence and Complex Systems, College of Engineering, Peking University, Beijing 100871, P. R. China}

\address[myfirstaddress]{Physics of Fluids Group, Max Planck Center Twente for Complex Fluid Dynamics, University of Twente, 7500 AE Enschede, The Netherlands}

\begin{abstract}
In a recent paper, Liu et al. [``Lift and drag in three-dimensional steady viscous and compressible flow'', Phys. Fluids \textbf{29}, 116105 (2017)] obtained a universal theory for the aerodynamic force on a body in three-dimensional steady flow, effective from incompressible all the way to supersonic regimes. In this theory, the total aerodynamic force can be determined solely with the vorticity distribution on a single wake plane locating in the steady linear far field. Despite the vital importance of this result, its validity and performance in practice has not been investigated yet. In this paper, we performed Reynolds-averaged Navier-Stokes simulations of subsonic, transonic, and supersonic flows over a three-dimensional wing. The aerodynamic forces obtained from the universal force theory are compared with that from the standard wall-stress integrals. The agreement between these two formulas confirms for the first time the validity of the theory in three-dimensional steady viscous and compressible flow. The good performance of the universal formula is mainly due to the fact that the turbulent viscosity in the wake is much larger than the molecular viscosity therein, which can reduce significantly the distance of the steady linear far field from the body. To further confirm the correctness of the theory, comparisons are made for the flow structures on the wake plane obtained from the analytical results and numerical simulations. The underlying physics relevant to the universality of the theory is explained by identifying different sources of vorticity in the wake. 
\end{abstract}


\end{frontmatter}

%\linenumbers

\section{Introduction}

In this paper we continue work from a previous paper of \citet{Liu2017-3d}. In that paper, \citet{Liu2017-3d} proved theoretically that, for the three-dimensional steady flow of viscous and compressible fluid, the aerodynamic force exerted on the body can be asymptotically expressed by vorticity integrals, 
\beq\lb{F-3D}
\pF = \rho_\infty \pU \times \int_{W} \px (\po \cdot \pe_x) \rd S + \fr{1}{2} \rho_\infty \pU \int_W \pe_x \cdot (\po \times \px) \rd S.
\eeq
Here $\pF$ is the total force, $\rho_\infty$ and $\pU$ are the density and velocity, respectively, of fluid at infinity, $\po$ is the vorticity, $\px$ is the coordinate vector, $W$ is the wake plane, and $\pe_x$ is the unit normal vector of $W$ which is assumed to be along the $x$-axis and parallel to $\pU$. An evident advantage of Eq.~\er{F-3D} is that the force is determined only by the vorticity moment on the wake plane, which is very convenient for experimental measurements and numerical simulations. The only limitation is that it may require a very large steady flow region such that a linear steady far field exists. This limitation, however, may be not so severe since most flows around commercial aircraft are nearly attached and fully turbulent. Despite the evident convenience of the universal force formula \er{F-3D}, its validity and performance in practice has not been explored yet. Thus, the first goal of the present paper is to assess the performance of Eq.~\er{F-3D} by performing numerical simulations. 

The universal force theory \er{F-3D}, however, reveals that no matter how many interacting processes could appear in a non-linear complex near-field flow, only the vorticity field has the farthest downstream extension such that its distribution can faithfully capture the total force. With no doubt, the information of total force must be included both in the transverse field and longitudinal field, but it seems only feasible to be extracted out from the former if only one field is involved. Actually, \citet{Wu1993} proved that, at least in the framework of the derivative moment transformation, the aerodynamic force cannot be determined solely with longitudinal variables. One of physical explanations for this phenomenon is that the transverse field is compact while the longitudinal field is dispersive. Since the longitudinal field makes a significant contribution to the aerodynamic force in the near-field flow while disappears in the far-field flow, there must be some mechanisms that can transform the information from the longitudinal field to the transverse field as the downstream location increases. Thus, the second goal of the present paper is to identity qualitatively and quantitatively the underlying mechanisms. 

The organization of this paper is as follows. In Section~\ref{sec.theory} a brief review of the universal force theory proposed by \citet{Liu2017-3d} is given, with emphasis on the underlying physics and assumptions. In Section~\ref{sec.numerics} the numerical simulations are reported for the subsonic, transonic, and supersonic flows over a three-dimensional wing. Then, the aerodynamic forces obtained from the universal formula \er{F-3D} and the wall-stress integral (see Eq.~\er{Fdef} below) are compared, aiming to test the validity and performance of the former. To further confirm the correctness of the theory, the flow structures on the wake plane are also studied by the analytical solutions and numerical simulations. In Section~\ref{sec.physics} the different sources of vorticity in the wake are identified, which provides a physical explanation for the universality of the force formula \er{F-3D}. Finally, the main findings of this work are summarized in Section \ref{sec.conclusion}.

\section{Universal force theory}\lb{sec.theory}

To obtain a universal force theory, one has to express the total force solely by kinematic variables in a form that the relevant boundary integrals of these variables are independent of the arbitrarily chosen boundary, such that the force formulae established by far-field linearized Navier-Stokes equations can well be applied to any boundary surrounding the body even in a highly non-linear flow zone. In general, this is possible only for steady flow where the force can be expressed by boundary integrals alone. A typical example is the classic circulation theorem for lift derived by \citet{Kutta1902} and \citet{Joukowski1906} (rewritten by \citet[][p.~404]{Batchelor1967}) and inflow theorem for drag derived by \citet{Filon1926}, where the circulation and inflow of incompressible flow are expressed by the boundary integrals of velocity potential and stream function, respectively. Of course the values of circulation and inflow depend on specific flow conditions and body geometry; but the force formulas remain universal and the existence of lift and drag depends only on the multi-valueness and/or singularity of the fields of velocity potential and stream function. 

In this section we highlight the universal force theory proposed by \citet{Liu2017-3d}, which extended the above idea to compressible flows. In this theory, the solid body is assumed to move steadily through the physical space filled with viscous and compressible fluid which is otherwise at rest. For convenience, the reference frame is fixed on the body such that a steady or statistically steady subspace $V_{\rm st}$ could exist before the flow reaches the truly unsteady far field \citep{Liu2017}. The dynamic viscosity $\mu$ is always assumed to be constant. Another implicit assumption is that the steady region $V_{\rm st}$ is sufficient large such that the flow in its far field can be linearized.

Let $\pu$ and $\pU$ denote the local and incoming flow velocities, then the disturbance velocity $\pu'=\pu - \pU$ can be written as
\beq\lb{HD}
\pu'=  \pu_\phi+\pu_\psi \equiv\na\phi+\na\times \ppsi, \quad \na\cdot\ppsi=0,
\eeq
where $\pu_\phi$ and $\pu_\psi$ are the longitudinal and transverse velocities, $\phi$ and $\ppsi$ are the velocity potential of the longitudinal field and the stream function of the transverse field, respectively. Then, the linearized compressible Navier-Stokes equation can be split as follows: 
\beqn
\Pi + \rho_\infty \pU \cdot \na \phi &\!\!\!=\!\!\!& 0, \lb{u-l} \\
\rho_\infty \pU \times \pu_\psi - \rho_\infty \na (\pU \cdot \ppsi) &\!\!\!=\!\!\!& \mu \po, \lb{u-t}
\eeqn
where $\Pi$ is the revised normal stress and $\mu$ is the dynamic viscosity. For steady flow, the total force $\pF$ exerted on the body $B$ can be transformed from the standard wall-stress integral
\beq\lb{Fdef}
\pF = -\int_{\pat B}(-\Pi \pn + \mu \po \times \pn)\textrm{d}S,
\eeq
where $\pat B$ is the surface of the body $B$ and $\pn$ is the corresponding unit normal vector, into the control surface integral, 
\beq\label{11.1b}
\pF = -\int_S (\Pi \pn +\rho \pu \pu\cdot \pn- \mu \po \times\pn) \mathrm{d} S, 
\eeq
where $S$ is an arbitrary control surface enclosing $B$ and locates inside $V_{\rm st}$, and $\pn$ is the unit outward normal vector of $S$. Hereafter we assume $S$ lies in sufficiently far away where the flow can be linearized and is thus governed by Eqs.~\er{u-l} and \er{u-t}. Using the exact continuity equation $\na\cdot (\rho \pu)=0$ and linearizing the integrands in Eq.~\er{11.1b}, there is 
\beq\lb{F2D}
\pF = \rho_\infty \pU\times \pGamma_\phi + \rho_\infty \pU Q_\psi + \rho_\infty \int_S \pn \times \na (\pU \cdot \ppsi) \mathrm{d} S,
\eeq
where 
\begin{equation}\label{G-phi}
\pGamma_\phi \equiv \int_S \pn \times \na \phi \mathrm{d} S, \quad
Q_\psi \equiv -\int_S \pn \cdot (\na \times \ppsi) \mathrm{d} S 
\end{equation}
are the circulation of longitudinal velocity and inflow of transverse velocity, respectively. In three dimensions the third term of Eq.~\er{F2D} has been proved to be the same as the first term \citep{Liu2017-3d}. Therefore, Eq.~\er{F2D} reduces to  
\beq\lb{f-uni}
\pF = 2 \rho_\infty \pU \times \pGamma_\phi + \rho_\infty \pU Q_\psi.
\eeq
According to the generalized Stokes theorem, $\phi$ and $\ppsi$ must be either multi-valued or singular otherwise there would be no force at all. Since this multi-valueness or singularity is independent of $S$, and hence so is Eq.~\er{f-uni}.

It should be noticed that, however, the universality and exactness of Eq.~\er{f-uni} are at the expense that $\pGamma_\phi$ and $Q_\psi$ cannot be measured directly in both experiments and simulations. To make up this disadvantage, one needs to find the circumstances in which these integrands can be replaced by physically observable variables. Therefore, we rewrite Eq.~\er{11.1b} as
\beq \lb{F2}
\pF= \rho_\infty \pU\times \pGamma - \rho_\infty \pU\cdot \int_S \pu_\psi \pn \mathrm{d} S +\mu \int_S \po\times \pn \mathrm{d} S, 
\eeq 
where 
\beq\label{eq.Gamma}
\pGamma \equiv \int_S \pn \times \pu \mathrm{d} S = \int_{V_{\rm st}} \po \mathrm{d} V = \int_{W} \px (\po \cdot \pe_x) \rd S
\eeq
is the circulation of total velocity, which is directly measurable. Note that the downstream face of $S$ is assumed to be the wake plane $W$. Since $S$ locates in the linear far field, the last term in Eq.~\er{F2} can be neglected, while the second term can be approximated as  
\beq \lb{F3}
- \rho_\infty \pU \cdot \int_S \pu_{\psi }\pn \mathrm{d} S = \fr{1}{2} \rho_\infty \pU \int_W \pe_x \cdot (\po \times \px) \rd S.
\eeq
Therefore, the universal force theory \er{F-3D} follows immediately.

\begin{figure*}[!tb]
\centering
\begin{overpic}[width=0.45\textwidth]{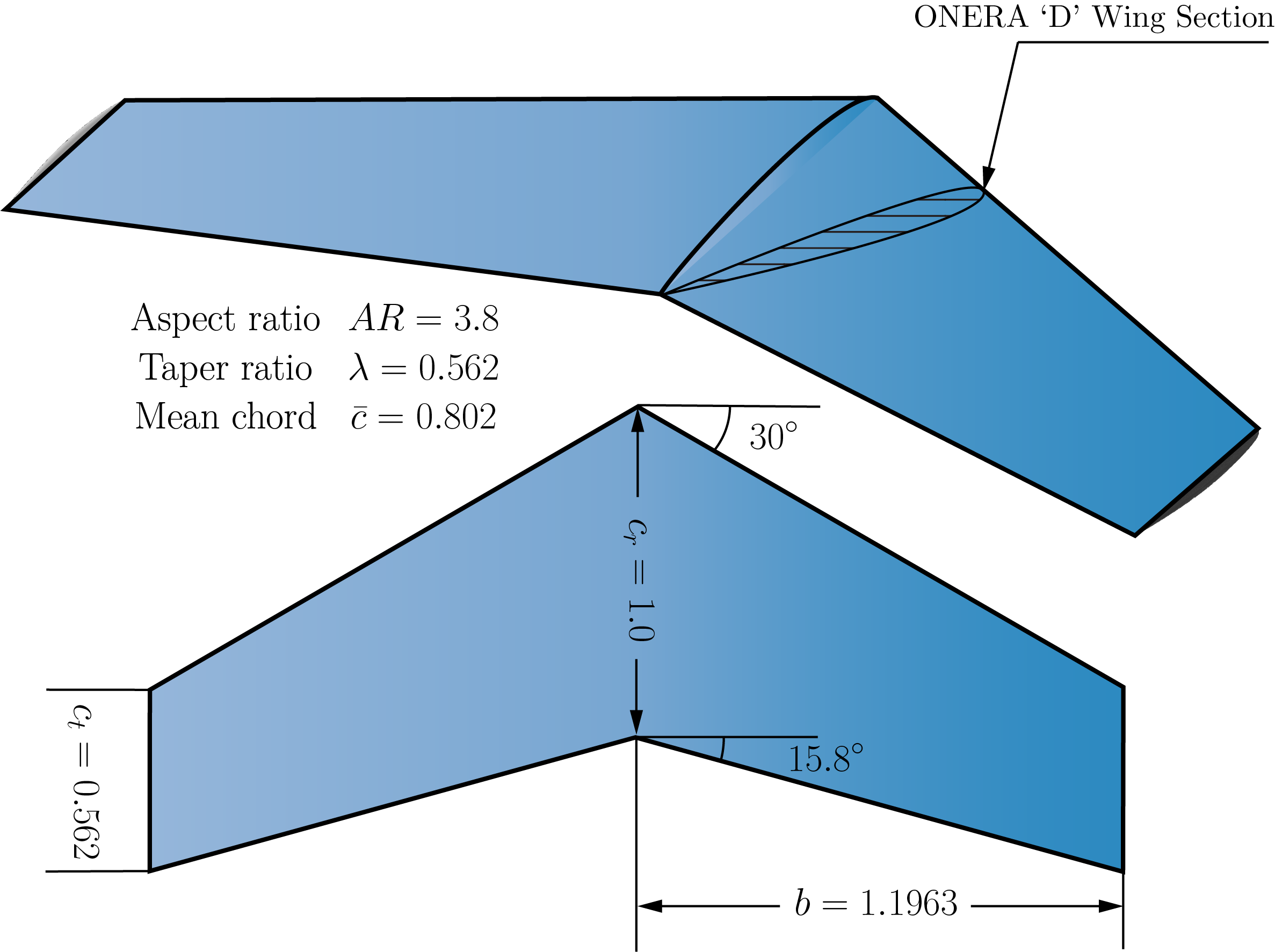}
\put(0,72){\footnotesize $(a)$}
\end{overpic}
\hspace{2mm}
\begin{overpic}[width=0.45\textwidth]{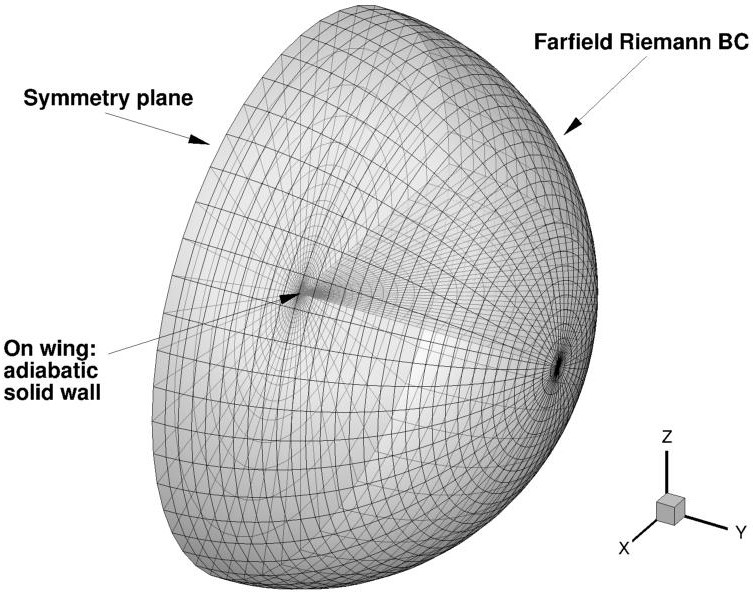}
\put(0,72){\footnotesize $(b)$}
\end{overpic}
\caption{(a) The geometry of the ONERA M6 wing. (b) The computational mesh.}
\label{fig.geometry}
\end{figure*}

Evidently, the validity of Eq.~\er{F-3D} relies heavily on the existence of a steady linear far field. Although the linear far field always exists, a steady subspace can be either present or absent depending on both flow conditions and body geometry. Even if a steady subspace exists, it still does not guarantee that a steady linear far field exists. Here we do not try to estimate the size of steady subspace. Instead, we always assume there is a steady subspace, of which the size is large enough to be regarded as an infinity space. Then, based on the solution of linearized compressible Navier-Stokes equation, one can give an estimation of the location of the steady linear far field \citep{Liu2017-3d}. For example, the downstream location of the three-dimensional linear far field is $x_m/c = O(C_D Re / 8 \pi)$, where $x_m$ is the minimum downstream location of the linear far field, $c$ is the characteristic length of the body (e.g., the root chord of a wing), $C_D = 2D/(\rho_\infty U^2 c^2)$ is the drag coefficient, and $Re=\rho_\infty U c/\mu$ is the Reynolds number. This estimation, although may be not accurate, provides the first condition under which good performance of Eq.~\er{F-3D} can be achieved.

\section{Numerical results}\lb{sec.numerics}

From the previous arguments we know that the universal force formula \er{F-3D} is valid from incompressible flow up to supersonic flow and its determination only requires the vorticity distribution on a single wake plane. These two characteristics make Eq.~\er{F-3D} be a breakthrough of classic aerodynamic force theory. However, its validity and performance has never been studied yet in practice. In this section, we will compare the forces obtained from the standard wall-stress integral \er{Fdef} and the universal force formula \er{F-3D} by performing numerical simulations of subsonic, transonic, and supersonic flows over a three-dimensional wing. To further confirm the correctness of the universal force theory, the flow structures on the wake plane are also studied both theoretically and numerically.

\subsection{Numerical method and validation}

The Stanford University Unstructured (SU$^2$) program developed by \citet{Palacios2013} is employed to solve the compressible Reynolds-averaged Navier-Stokes (RANS) equations. In this open-source program, the RANS equations are solved by a second-order accurate finite-volume method, with the convective terms discretized by the Roe scheme for shock capture and the viscous terms by a least-squares method. For temporal terms an implicit Euler scheme is adopted. The Spalart-Allmaras turbulence model is employed for the closure of RANS equations. The SU$^2$ program has been used widely and verified for different cases, see, for example, \citet{Palacios2013}.

The three-dimensional wing developed by the Office National d'Etudes et de Recherches A\'erospatiales (ONERA) is adopted here to investigate the validity and performance of the universal force formula \er{F-3D}, as well as the flow topological structures on the wake plane locating in the linear far field. Figure~\ref{fig.geometry}(a) shows the geometrical parameters of the ONERA M6 wing. It is a swept wing with no twist and uses a symmetric airfoil of the ONERA `D' wing section. The ONERA M6 wing is a classic numerical simulation validation case for external flows because of its simple geometry combined with complexities of transonic flow. In order to capture the flow structures in near field and far field accurately at the same time, the standard computational results for the ONERA M6 wing provided by the CFL3D (Computational Fluids Laboratory 3-Dimensional) code in near field are served as the initial value for the SU$^2$ code. This approach significantly improves the prediction of shock wave location on the upper surface and the flow topological structures in the far field.  

Figure~\ref{fig.geometry}(b) shows the standard computational mesh for the flow over the ONERA M6 wing. It is a O-H type structured grid, of which the outer boundary is a spherical surface of radius $r=100c$, where $c$ is the root chord of the wing (see Fig.~\ref{fig.geometry}a). The orthogonality of the grid is very good and the grid resolution in the wake region is much higher aiming to capture the flow topological structures on the wake plane accurately. The grid independence has been confirmed by \citet{Gao2019}, where the results obtained from both the coarse grid with 1.08 million total grid number and fine grid with 8.64 million total grid number agree very good with the wind tunnel measurements. In the remaining of the paper, the data obtained from the same fine grid is selected for analyses.

\subsection{Lift and drag}

\begin{figure*}[!t]
\centering
\begin{overpic}[width=0.32\textwidth]{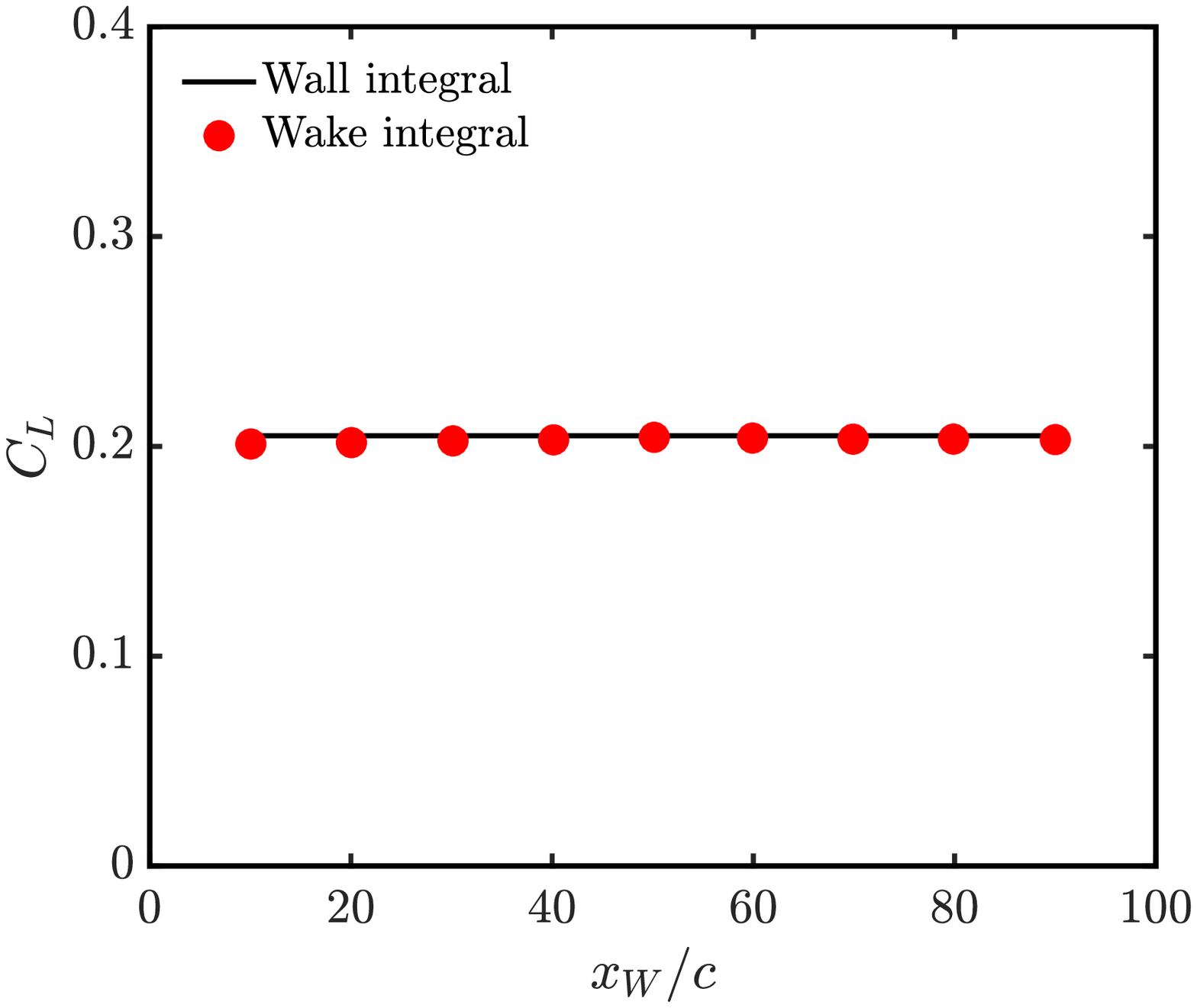}
\put(0,74){\footnotesize $(a)$}
\end{overpic}
\begin{overpic}[width=0.32\textwidth]{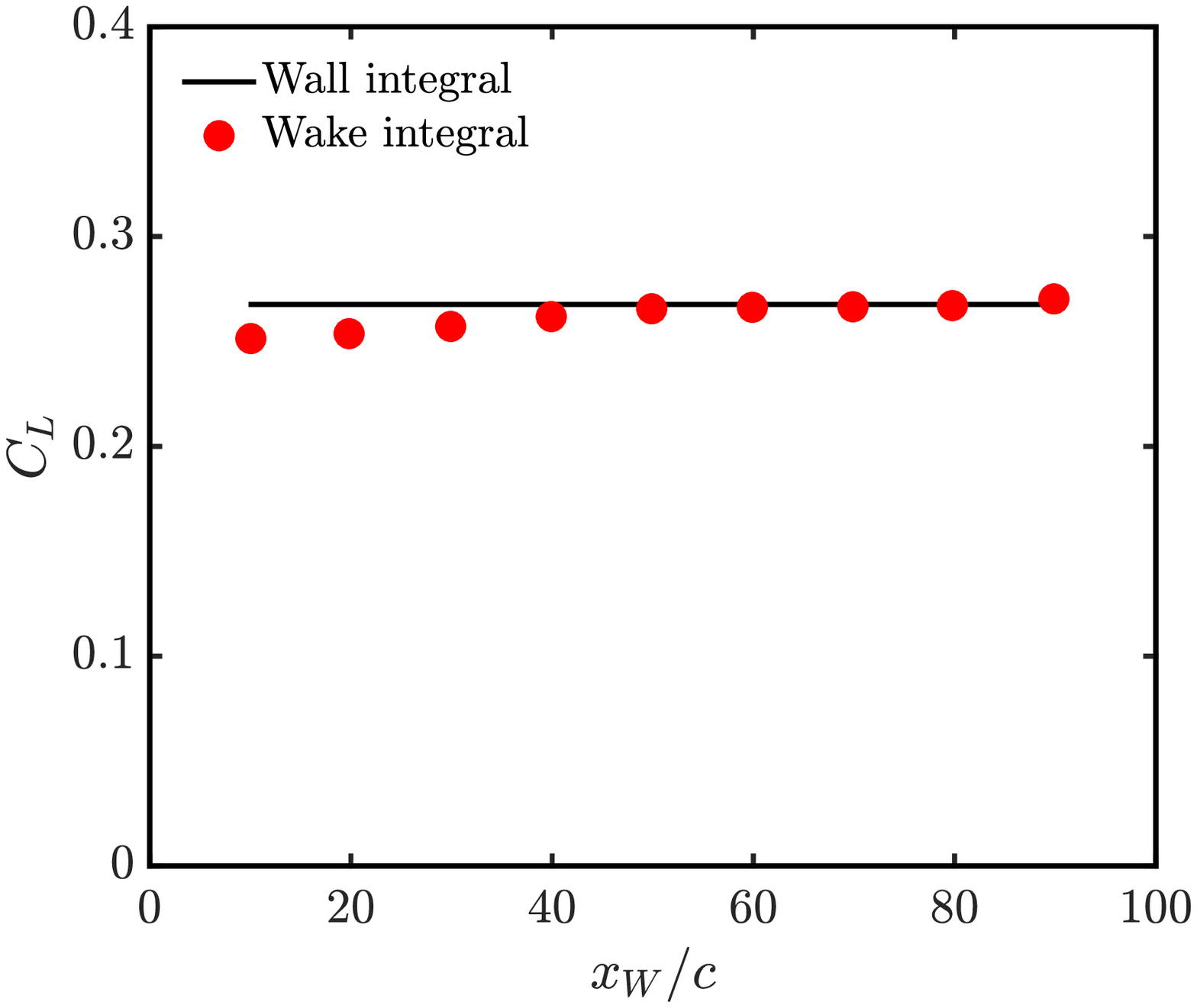}
\put(0,74){\footnotesize $(b)$}
\end{overpic}
\begin{overpic}[width=0.32\textwidth]{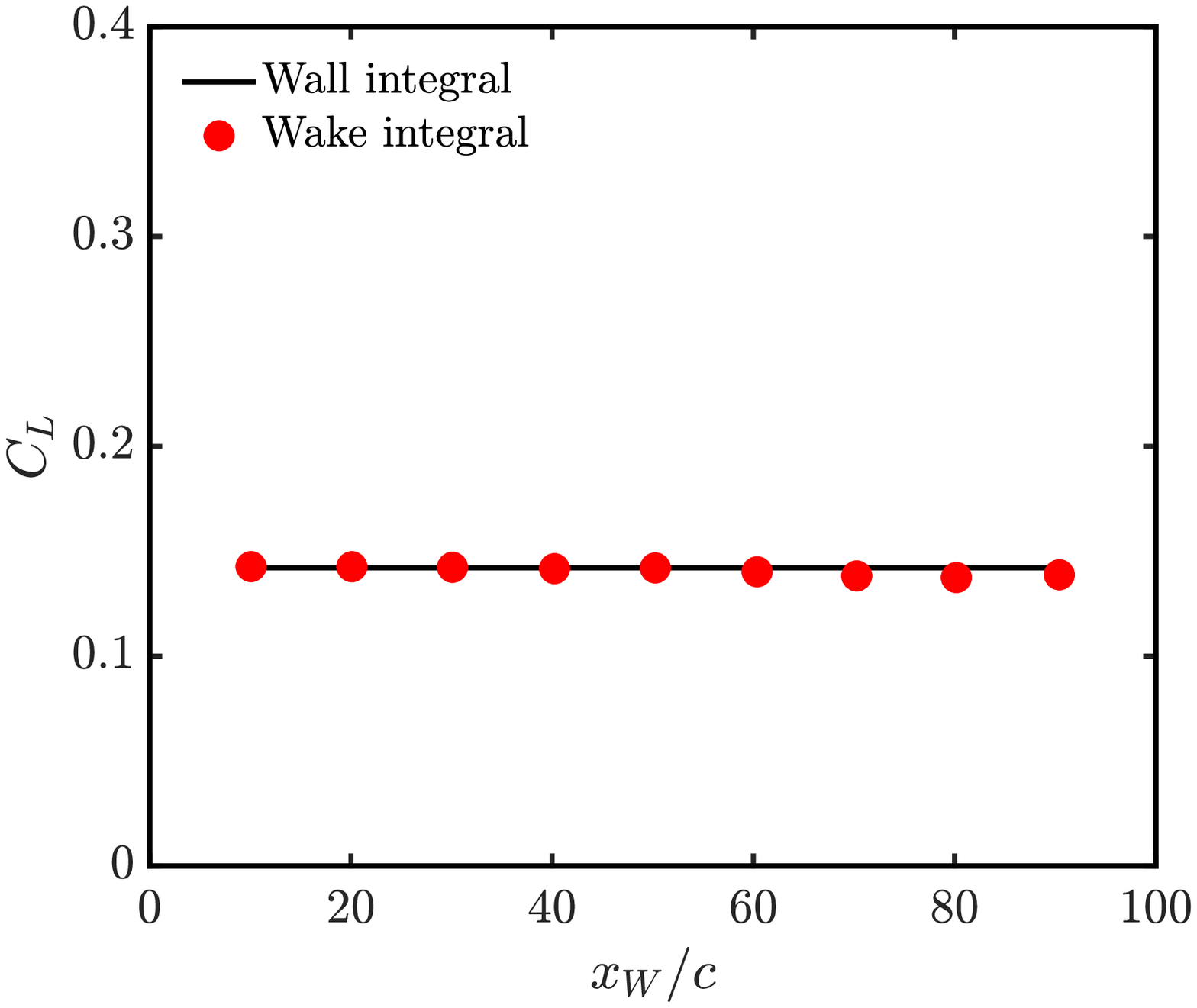}
\put(0,74){\footnotesize $(c)$}
\end{overpic}
\begin{overpic}[width=0.32\textwidth]{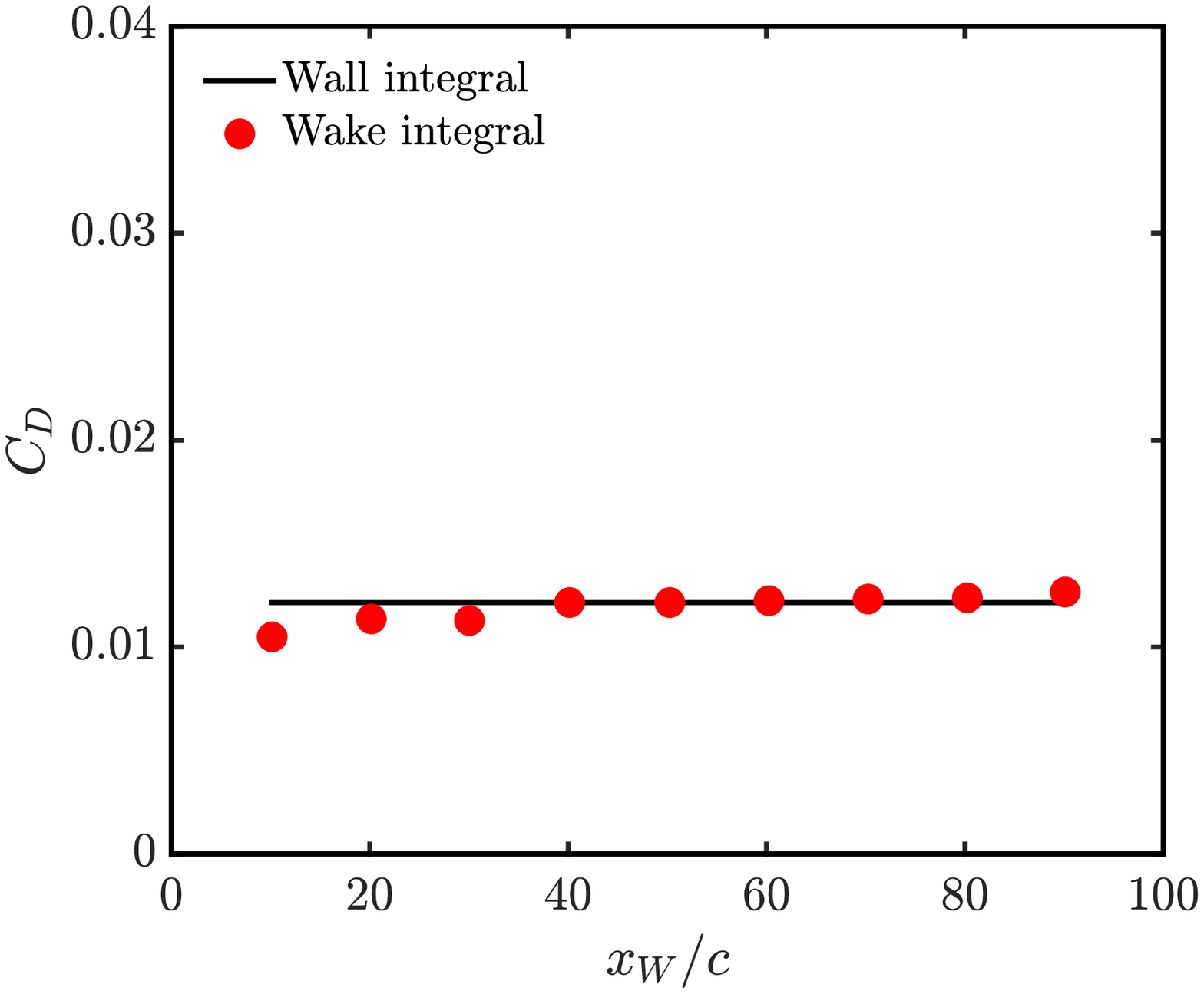}
%\put(0,118){\footnotesize $(a)$}
\end{overpic}
\begin{overpic}[width=0.32\textwidth]{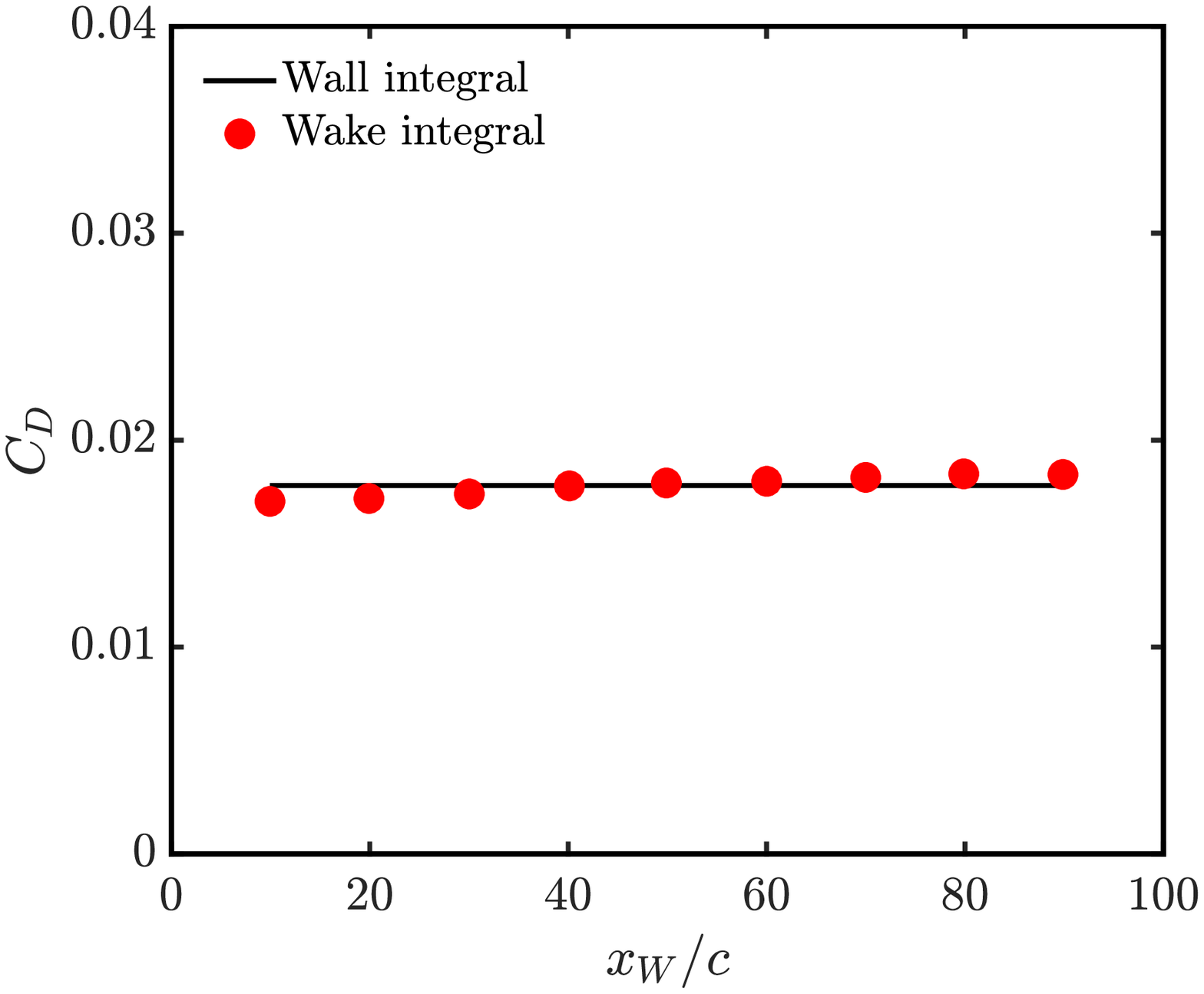}
%\put(0,118){\footnotesize $(a)$}
\end{overpic}
\begin{overpic}[width=0.32\textwidth]{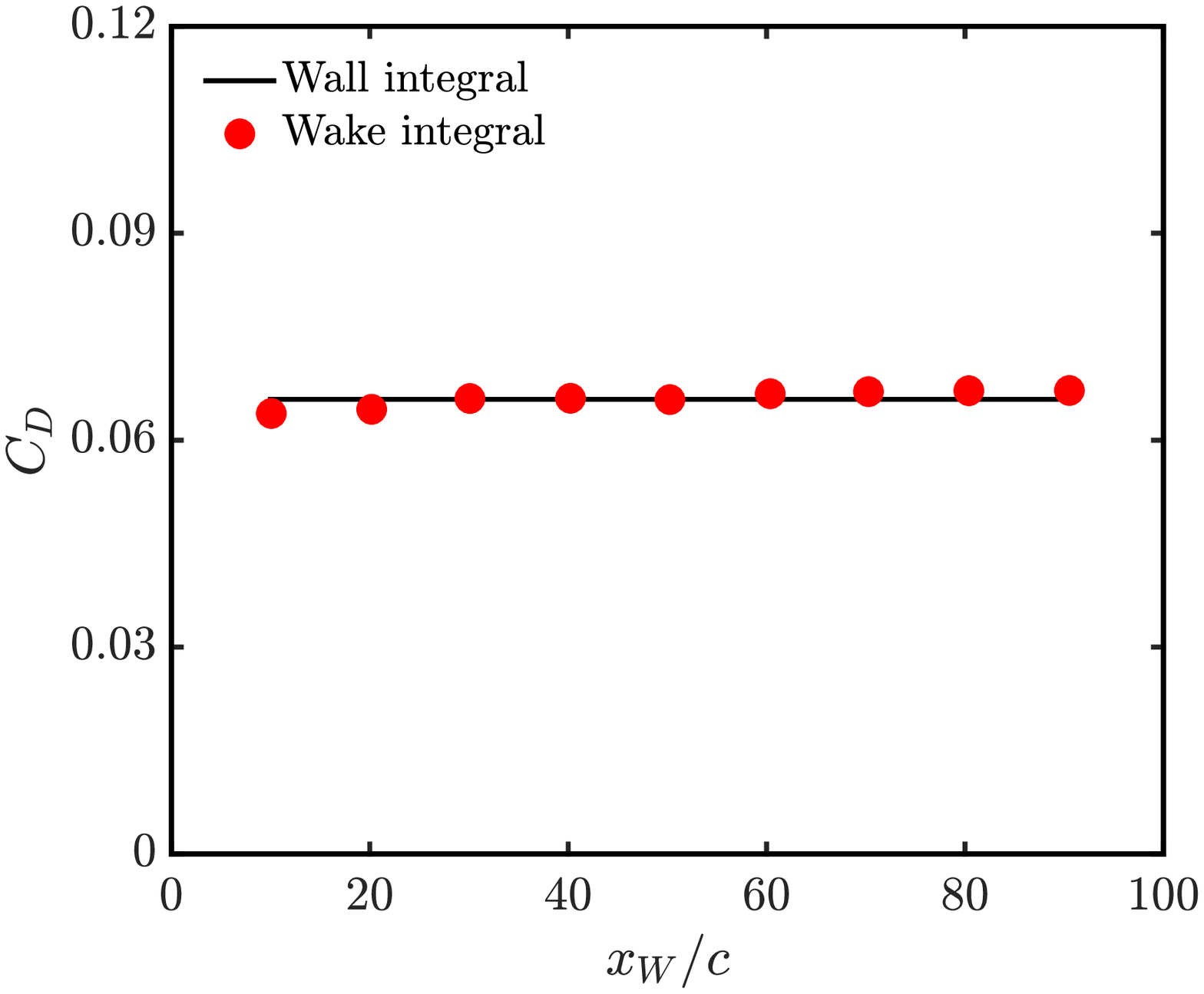}
%\put(0,118){\footnotesize $(a)$}
\end{overpic}
\caption{The lift and drag coefficients of the ONERA M6 wing as a function of downstream location of the wake plane. (a) $M_\infty=0.5$, (b) $M_\infty=0.8395$, and (c) $M_\infty=1.6$. Solid lines: results of wall-stress integral \er{Fdef}; Symbols: results of wake plane integral \er{F-3D}. }
\label{fig.lift-drag}
\end{figure*}

To investigate the validity and performance of the universal force formula \er{F-3D}, we performed three simulations with incoming Mach number $M_\infty = 0.5$, $0.8395$, and $1.6$, which corresponds to subsonic, transonic, and supersonic flows, respectively. All the other parameters are the same as that in the previous subsection.

Figure~\ref{fig.lift-drag} shows the comparisons of the lift coefficient, $C_L = L/(\rho_\infty U^2 b c)$, and drag coefficient, $C_D = D/(\rho_\infty U^2 b c)$, calculated from the wall-stress integral \er{Fdef} and the wake plane integral \er{F-3D} as the wake plane locates at different downstream positions, i.e., $x/c \in [10, 90]$ with the origin locating at the leading edge of the wing. In these simulations, the lift is about one order larger than the drag. Thus, the former can be predicted more easily and accurately than the latter. Overall, the results obtained from the wake integral agree very well with that from the wall integral for all tested Mach numbers and downstream locations. In particular, all maximum relative errors for both the lift and drag are less than 1\% as long as $x/c \ge 40$, which can be approximated as the minimum downstream location of the steady linear far field. This excellent agreement confirms the prediction of the universal force formula \er{F-3D}. We also notice that the performance of Eq.~\er{F-3D} becomes better and better as the wake plane moves further downstream provided that the grid resolution is still high enough therein and the wake plane is not too close to the downstream boundary.

\subsection{Downstream location of the linear far field}

As remarked at the end of Section~\ref{sec.theory}, the downstream location of the linear far field is $x_m/c = O(C_D Re/8\pi)$, which for the case $M_\infty=0.5$ should be $x_m/c = O(10^3)$. However, results shown in Fig.~\ref{fig.lift-drag} indicates that $x/c = O(10^1)$, much less than the theoretical estimation. This difference is mainly due to the laminar assumption in the theoretical estimation where the viscosity is assumed constant with the same value as that at infinity. In contrast, all simulated flows are turbulent, of which the turbulent viscosity is much larger than the molecular viscosity, especially in the wake region. Taking this fact into account could correct the estimation.

\begin{figure}[!tb]
\centering
\begin{overpic}[width=0.45\textwidth]{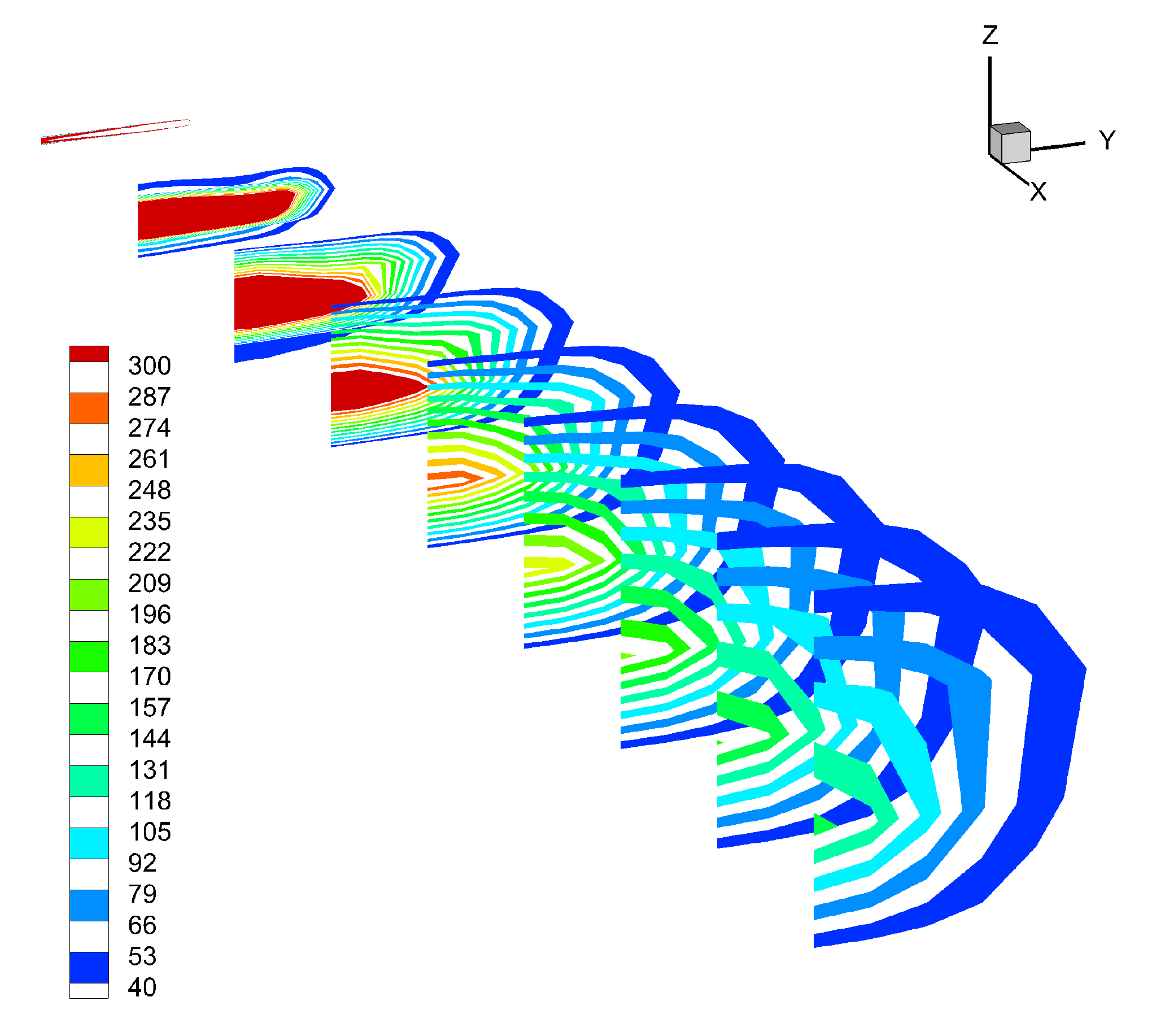}
%	\put(0,113){\footnotesize $(b)$}
\end{overpic}
\caption{The contour of the ratio between the turbulent viscosity and molecular viscosity with $M_\infty=0.5$ at downstream $x/c\in[1, 20]$.}
\label{fig.viscosity}
\end{figure}

\begin{figure*}[!t]
\centering
\begin{overpic}[width=0.38\textwidth]{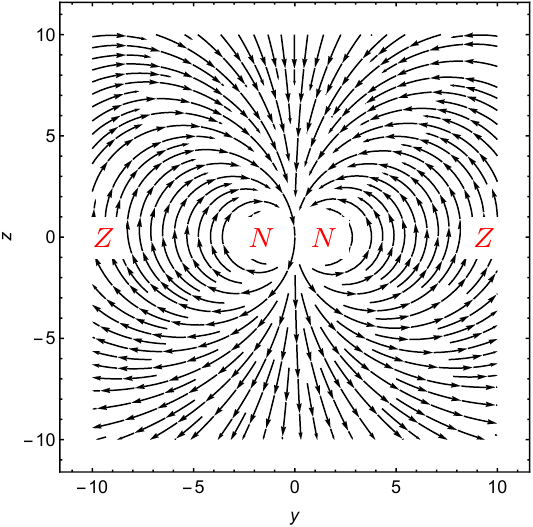}
\put(0,95){\footnotesize $(a)$}
\end{overpic}
\hspace{2mm}
\begin{overpic}[width=0.38\textwidth]{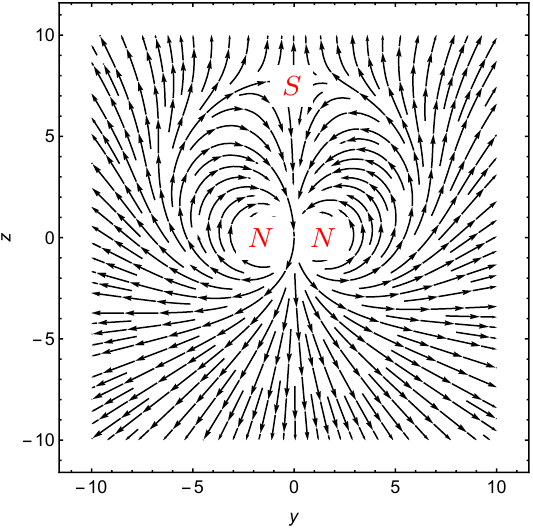}
\put(0,95){\footnotesize $(b)$}
\end{overpic}
\begin{overpic}[width=0.38\textwidth]{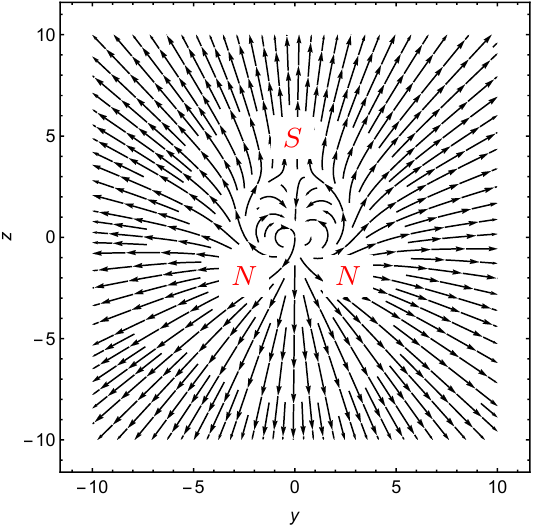}
\put(0,95){\footnotesize $(c)$}
\end{overpic}
\hspace{2mm}
\begin{overpic}[width=0.38\textwidth]{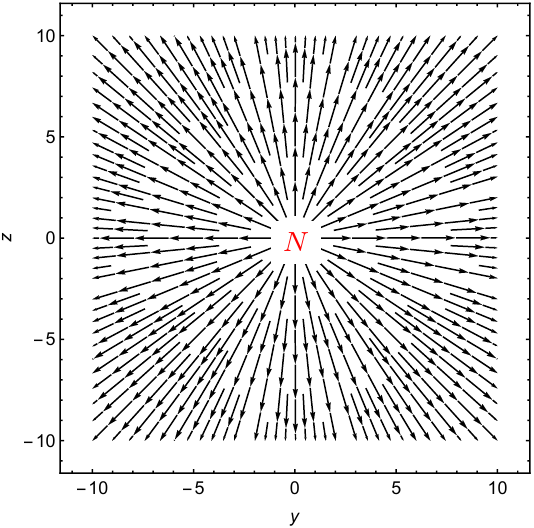}
\put(0,95){\footnotesize $(d)$}
\end{overpic}
\caption{Analytical flow topological structures on the wake plane locating in the linear far field with $M_\infty=0.5$, $k=10$, $x=10$, and $\pF = (D, 0, L)$. (a) $L/D=1$, (b) $L/D=0.1$, (c) $L/D=0.01$, and (d) $L/D=0$. }
\label{fig.subsonic}
\end{figure*}

Figure~\ref{fig.viscosity} shows the development of the ratio $\eta=\mu_t/\mu$ between turbulent viscosity $\mu_t$ and molecular viscosity $\mu$ in the wake region at the downstream location $x/c \in [1, 20]$, for the case $M_\infty=0.5$. It can be observed that, close to the wing the maximum value of $\eta$ is larger than 300, and at the downstream position $x/c=20$ the maximum value of $\eta$ is still larger than 100. Therefore, we can assume $\eta = O(10^2)$ such that the downstream location of the linear far field is $x_m/c = O(C_D Re/8\pi \eta) = O(10^1)$, which is now consistent with the results indicated by Fig.~\ref{fig.lift-drag}.

Although the downstream location of the linear far field cannot be determined exactly, its estimation is still very helpful since it gives a lower limit where Eq.~\er{F-3D} can predict the aerodynamic force accurately with only the information of the vorticity distribution on the wake plane, from incompressible all the way to the supersonic regimes.

\subsection{Topological structures of the flow on wake planes}

There are two purposes to study the topological structures of the flow on wake planes. First, it may be helpful to estimate the order of the lift-drag ratio. Second, it can be used to confirm the correctness of the numerical results. In general, the velocity vectors on an arbitrary finite region  $\mathcal{D}$ of a wake plane $W$ can move freely inward and outward across the boundary $\partial \mathcal{D}$, but the numbers of the isolated singular points (e.g., saddles and nodes) and the boundary switch points (the points on the boundary at which the vectors are tangential to the boundary segment therein) are conserved due to the Poincare–Bendixson index theorem \citep{Liu2011}, 
\beq\lb{eq.liu}
\sum_N - \sum_S = 1 + \frac{1}{2} \sum_{Z},
\eeq
where $\sum_N$ is the number of nodes, $\sum_S$ is the number of saddle points, and $\sum_{Z}$ is the number of switch points, respectively.

\begin{figure*}[!t]
\centering
\hspace{6mm}
\begin{overpic}[width=0.35\textwidth]{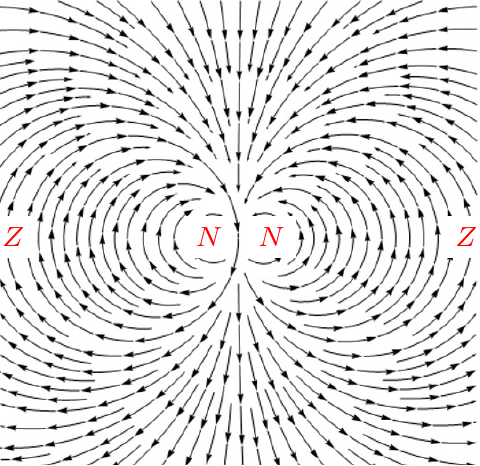}
\put(-8,93){\footnotesize $(a)$}
\end{overpic}
\hspace{12mm}
\begin{overpic}[width=0.35\textwidth]{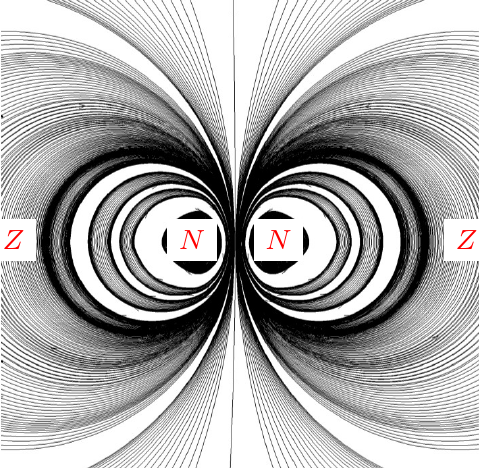}
\put(-8,93){\footnotesize $(b)$}
\end{overpic}
\caption{Flow topological structures on the wake plane with $M_\infty=0.8395$. (a) Analytical results and (b) numerical simulations.}
\label{fig.transonic}
\end{figure*}

\begin{figure*}[!t]
\centering
\hspace{6mm}
\begin{overpic}[width=0.35\textwidth]{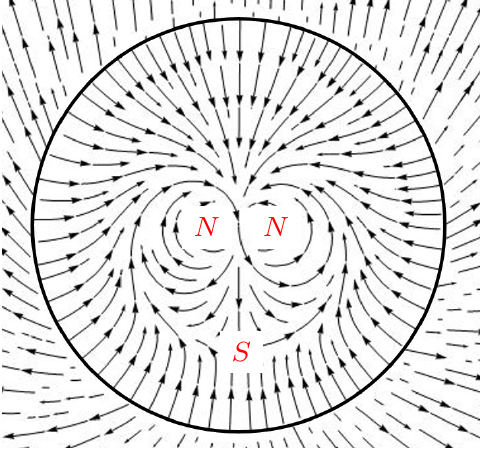}
\put(-8,90){\footnotesize $(a)$}
\end{overpic}
\hspace{12mm}
\begin{overpic}[width=0.35\textwidth]{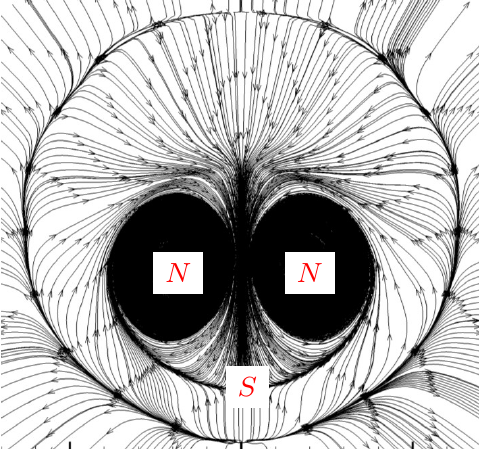}
\put(-8,90){\footnotesize $(b)$}
\end{overpic}
\caption{Flow topological structures on the wake plane with $M_\infty=1.6$. (a) Analytical results and (b) numerical simulations.}
\label{fig.supersonic}
\end{figure*}

Figure~\ref{fig.subsonic} shows the analytical streamlines on the wake plane $W$ located in the linear far field with $M_\infty=0.5$, $k=10$ and $x=10$. These streamlines are obtained from Eq.~\er{HD} with $\phi$ and $\ppsi$ given by \citet{Liu2017-3d}, 
\beqn \lb{eq.subsonic}
\phi &\!\!\!=\!\!\!& \frac{1}{4 \pi \rho_\infty U} \pF \cdot \na \ln (r_\beta - x), \\
\ppsi &\!\!\!=\!\!\!& -\frac{1}{4 \pi \rho_\infty U} \pF \times \na \Gamma[0, k(r-x)], 
\eeqn
where $r_\beta = \sqrt{x^2 + (1-M_\infty^2) (y^2+z^2)}$, $k=\rho_\infty U/(2\mu)$ and $\Gamma(\cdot, \cdot)$ is the upper incomplete gamma function. To study the dependence of the streamline pattern on the lift-drag ratio $L/D$, flow patterns obtained with $L/D=1, 0.1, 0.01, 0$ are displayed in the figure. When the velocity induced by lift dominates the flow, the flow pattern is the same as that induced by a pair of trailing vortex. Thus, there are two nodes $N$ and two switch points $Z$ in Fig.~\ref{fig.subsonic}(a). When the flow is induced only by the drag, the flow pattern is the same as that induced by a point source and thus only one node $N$ exists in Fig.~\ref{fig.subsonic}(d). When the flow is induced by both the lift and drag, a saddle point $S$ appears on the upper part of the wake plane, see Fig.~\ref{fig.subsonic}(b,c). As the lift reduces to zero, the saddle point moves close to the nodes and ultimately cancels one of the nodes (Fig.~\ref{fig.subsonic}d). Nevertheless, in all these cases the number of nodes, saddle points, and switch points satisfies Eq.~\eqref{eq.liu}.

Figures~\ref{fig.transonic} compares the flow topological structures, obtained from the analytical prediction and numerical simulation of flow over the ONERA M6 wing, on the wake plane locating in the linear far field with $M_\infty = 0.8395$. Qualitatively, these two results agree very well with each other, with two nodes caused by the trailing vortex pair and two switch points due to the shape of the finite section, which is almost the same as subsonic flow when $L/D$ is not too small (see Fig.~\ref{fig.subsonic}a). 

Figure~\ref{fig.supersonic} compares the corresponding flow topological structures on the wake plane in the linear far field with $M_\infty = 1.6$. Qualitatively, these two results also agree well with each other. In particular, outside the Mach cone, the disturbance velocity is induced by the viscous diffusion, of which the magnitude is extremely small. But the streamline pattern is similar to that caused by a single source. Inside the Mach cone, there are still two nodes caused by the trailing vortex pair and one saddle point due to the competing effect of the flow induced by the lift and drag. In contrast to subsonic flow with relatively small $L/D$ where the saddle point appears in the upper part of the wake plane (see Fig.~\ref{fig.subsonic}(b,c)), now the saddle point appears in the lower part. In addition, the streamline pattern is very similar to that caused by a single sink, especially for very small $L/D$ (results not shown here).

\section{Physical discussions}\lb{sec.physics}

It is well known that in the classic supersonic aerodynamic theory shock waves may affect the lift, and the drag can be divided into the friction drag, form drag, induced drag, and wave drag, etc. In viscous flow, the transverse field is directly characterized by vorticity and responsible for the friction drag and induced drag, while the longitudinal field is directly characterized by dilatation or thermodynamic variables and responsible for the wave drag. In the near field, these two fields are inherently coupled with each other. In particular, inside the flow the coupling happens via non-linearity, for example the generation of vorticity by curved shocks and that of shear layers by shock interactions. On the body surface, the coupling happens via viscosity and no-slip condition, for example vorticity generation by pressure gradient. Thus, the information of the aerodynamic force must be included both in the transverse field and longitudinal field. The universal force theory \er{F-3D}, however, reveals that the aerodynamic force can be determined solely by the vorticity distribution on the wake plane. On the one hand, since the information of total force included in the longitudinal field won't disappear automatically, there must be some physical mechanisms that can transform the information from the longitudinal field to the transverse field as the downstream location increases. On the other hand, because the total force can be determined solely by the vorticity distribution on the wake plane, these mechanisms must be related to the different sources of vorticity in the wake. 

\begin{figure}[!tb]
\centering
\hspace{-2mm}
\begin{overpic}[width=0.45\textwidth]{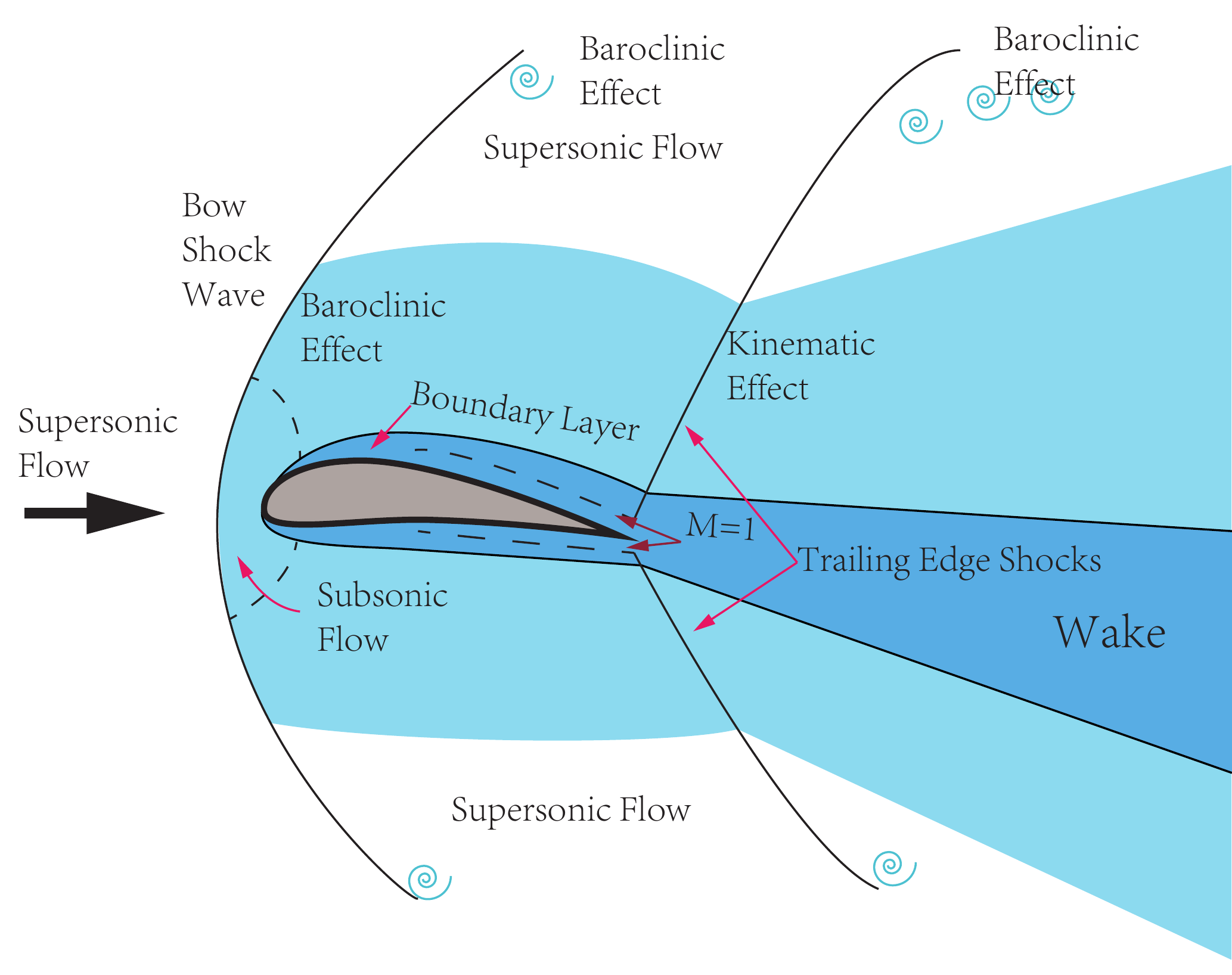}
%	\put(0,113){\footnotesize $(b)$}
\end{overpic}
\caption{Sketch of different sources of vorticity in the wake of supersonic flow over an airfoil}
\label{fig.super-wake}
\end{figure}

Figure~\ref{fig.super-wake} sketches out the different sources of vorticity in the wake of supersonic flow over an airfoil. To explain the underlying physics relevant to these sources, we first derived the vorticity jump behind a shock wave in steady flow (for details see Appendix \ref{sec.append}),
\beq \lb{eq.jump}
m [\! [ \po_\pi ]\! ] = [\![ \rho ]\!] \omega_n \pu_\pi + [\![ u_n ]\!] \pn \times \na m - [\![ \rho ]\!] u_\pi \pn \times \na u_\pi,
\eeq
where $[\![ \cdot ]\!]$ denotes the jump across the shock, subscripts $n$ and $\pi$ refer to the normal and tangential components, respectively, $\pn$ is the unit normal vector of the shock, and $m=\rho u_n$ is the mass flux. Equation~\er{eq.jump} reveals that there are two different mechanisms through which the shock waves can affect the generation of vorticity. The first one is the well-known baroclinic effect and governed by the last two terms in Eq.~\er{eq.jump}, which indicates that new vorticity can be generated through curved shocks even if the incoming flow is irrotational. The second one is the kinematic effect and governed by the first term in Eq.~\er{eq.jump}, which indicates that new vorticity can be generated through either curved or straight shocks as long as the normal vorticity component of the incoming flow is non-zero. These two effects are explicitly shown in the figure. In the far field, only the baroclinic effect is significant since the shock waves are too weak. In the near field, the dominant one is the baroclinic effect near the leading edge of the airfoil since the shock is curved and the incoming flow is nearly irrotational, while it is the kinematic effect near the trailing edge as the shock wave is nearly straight therein. In addition, the solid boundary provides an additional source of vorticity, which is a direct result of linear coupling between the transverse and longitudinal processes and is the only source of vorticity in incompressible flow. As shown in Fig.~\ref{fig.super-wake}, the boundary layer vorticity below the sonic line can be directly advected downstream to the far wake without the modulation of the shock, while that above the sonic line will interact with the trailing-edge shock. Nevertheless, it is via these coupling mechanisms that the information of longitudinal field is transformed into the vorticity field and hence enables one to determine the aerodynamic force solely by the latter.

We emphasis that, the underlying mechanism why the vorticity alone can capture faithfully the total aerodynamic force has already been indicated quantitatively in the derivation of the universal force theory \er{f-uni} from Eq.~\er{F2D}. In the derivation (for details please see \citet{Liu2017-3d}), only the fundamental solution of the transverse field has indeed been taken into account for the calculation of the aerodynamic force, while the fundamental solution of the longitudinal field that has different behavior under different Mach number does not appear. This is because the former alone is sufficient to represent the contribution of the multi-valueness and singularity to the aerodynamic force. However, the physical mechanism for the disappearing of the longitudinal field is still due to the coupling mechanisms between the transverse and longitudinal fields, although in the fundamental solution approach this coupling reduces to a Delta function.

\section{Conclusions}\lb{sec.conclusion} 

In this paper, we performed Reynolds-averaged Navier-Stokes simulations of subsonic, transonic, and supersonic turbulent flows over the ONERA M6 wing. The total aerodynamic forces are calculated with both the standard wall-stress integral and the testable universal force formula. The excellent agreement between the results obtained from these two formulas confirms numerically the validity of the universal force theory. Due to the turbulence effect, the downstream location of the linear far field is found much nearer than theoretical prediction, which leads to the good performance of the testable universal force formula in practice. In particular, the maximum relative errors of all tested cases are less than 1\% when $x/c \ge 40$. The flow topological structures on the wake plane are also studied numerically and theoretically, of which the qualitative agreement further confirms the correctness of the theory. Finally, the underlying physics relevant to the universality of the theory is explained by identifying different sources of vorticity in the wake.

%%%% Acknowledgments %%%%%%%%
\section*{Acknowledgments}
\addcontentsline{toc}{section}{Acknowledgements}

This work was supported by the National Natural Science Foundation of China (Grant No. 11472016). The authors are grateful to Drs. Ankang Gao and Linlin Kang for valuable discussions. Our special thanks go to Prof. Cunbiao Lee because the simulations were performed at his workstation.

\appendix
\addcontentsline{toc}{section}{Appendix}

\section{Vorticity jump across a steady shock}\lb{sec.append}

It is well known that the general expression for the vorticity jump across a two-dimensional inviscid steady shock in a uniform flow is first obtained by \citet{Truesdell1952}, 
\beq \label{eq:truesdell}
[\![ \om ]\!] = - \frac{(1- \epsilon)^2}{\epsilon} u_s K, \quad \epsilon = \frac{\rho_0}{\rho_1} \le 1,
\eeq
where $[\![ \cdot ]\!] = (\cdot)_1 - (\cdot)_0$, with subscripts $0$ and $1$ referring to quantities ahead of and behind the shock, respectively; $u_s$ and $K$ are tangential component of velocity and curvature, respectively. The corresponding three-dimensional theory was developed by \citet{Lighthill1957} and \citet{Hayes1957}. In terms of the intrinsic streamline coordinates, \citet{Hayes1957} obtained the expression valid in both the uniform and non-uniform flows:
\beq \label{eq:hayes-s}
[\![ \po_\pi ]\!] = \pn \times \left( \nabla m \left[\!\!\left[ \frac{1}{\rho} \right]\!\!\right] - \frac{[\![ \rho ]\!]}{m} \pu_\pi \cdot \nabla \pu_\pi \right).
\eeq
However, the relation between shock wave and vorticity jump cannot be explained by term $\pu_\pi \cdot \nabla \pu_\pi$ intuitively and clearly. Based on that, \citet{Kevlahan1996, Kevlahan1997} expanded this term in two dimensions and used it to explain the vorticity generated by shocklets in compressible turbulent flow. Following the work of \citet{Lighthill1957} and \citet{Hayes1957}, \citet{Wu2006} proposes a simple and general way to re-derive vorticity jump condition in three-dimensional general steady flow (with small error), which reveals the physical mechanisms behind in vorticity across the shock wave clearly. Now we follow the same way and re-derive it below.

Take the jump of the tangential component of the steady Euler equation:
\beqn\label{eq:euler-s1}
& & \pn \times [\![ \rho (\po \times \pu + q\nabla q) ]\!] = -\pn \times [\![ \nabla p ]\!], \\ 
& & q^2 = |\pu|^2 =u_n^2 + u_\pi^2.\label{eq:euler-s2}
\eeqn
In this equation there are jumps of some products, say $[\![ fg ]\!]$, which can be treated by using a pair of identities:
\beq
[\![ f  g ]\!] = \bar{f} [\![ g ]\!] + [\![ f ]\!]\bar{g}, \quad
\overline{fg} = \bar{f} \bar{g} + \frac{1}{4} [\![ f ]\!] [\![ g ]\!],
\eeq
where $\bar{f} = (f_1+f_0)/2$. Taking Rankine-Hugoniot shock relations into consideration, we obtain:
\beqn 
\pn \times [\![ \rho (\po \times \pu) ]\!] &\!\!\!=\!\!\!& m[\![ \po_\pi ]\!]-[\![ \rho  ]\!]\pu_\pi \om_n, \label{eq:t1} \\
\pn \times [\![ \nabla p+ \rho q \nabla q ]\!] &\!\!\!=\!\!\!& [\![ \rho ]\!]u_\pi \pn \times \nabla u_\pi - [\![ u_n ]\!] \pn \times \nabla m \label{eq:t2}.
\eeqn 
Thus, from Eqs.~\er{eq:euler-s1} and \er{eq:euler-s2} we obtain a general formula for the vorticity jump behind a shock in steady flow:
\beq\label{eq:om-j}
m [\! [ \po_\pi ]\! ] = [\![ \rho ]\!] \omega_n \pu_\pi + [\![ u_n ]\!] \pn \times \na m - [\![ \rho ]\!] u_\pi \pn \times \na u_\pi.
\eeq
It is straightforward to verify that Eqs.~\eqref{eq:om-j} and \eqref{eq:hayes-s} are equivalent.

%\section*{References}
\addcontentsline{toc}{section}{References}
\bibliographystyle{model5-names}\biboptions{authoryear}
\bibliography{mybibfile}

\end{document}